\documentclass[12pt]{article}
\begin{document}

\begin{center}
{\LARGE \bf Position-dependent mass models and their nonlinear
characterization}\\
\vspace{2cm}
{\bf B Bagchi $^{^{\ast}}$}\\
\vspace{.5cm}
{\small Department of Applied Mathematics\\
University of Calcutta\\
92 Acharya Prafulla Chandra Road\\
Kolkata - 700 009}\\

\vspace{3cm} {\bf Abstract}
\end{center}

\indent We consider the specific models of Zhu-Kroemer and
BenDaniel-Duke in a sech$^{2}$-mass background and point out
interesting correspondences with the stationary 1-soliton and
2-soliton solutions of the KdV equation in a supersymmetric
framework.
 \\\\\\\\\\\\\\\\


--------------------
\\$\ast$ E-mail: bbagchi123@rediffmail.com
\\

\newpage

\indent  In dealing with position dependent mass (PDM) models
controlled by a sech$^{2}$-mass profile, we demonstrated [1]
recently that, in the framework of a first-order intertwining
relationship, such a mass environment generates an infinite
sequence of bound states for the conventional free-particle
problem. Noting that the intertwining relationships are naturally
embedded in the formalism [2] of the so-called supersymmetric
quantum mechanics (SUSYQM), we feel tempted to dig this issue a
little deeper by choosing to examine the connections between the
discrete eigenvalues of such a PDM quantum Hamiltonian
(transformed appropriately so that a SUSY structure is evident)
and the stationary soliton solutions of the Korteweg de Vries
(KdV) equation that match with the mass function upto a
constant of proportionality.\\

Let us begin with the standard  time-independent representation of
the PDM Schr\"{o}dinger equation [3]
$$\Big[-\frac{d^{2}}{dx^{2}}+\frac{3}{4}\frac{M'^{2}}{M^{2}}-\frac{1}{2}\frac{M''}{M}+M(V_{eff}-\epsilon)\Big]\psi=0\eqno(1)$$where
$M(x)$ is the dimensionless equivalence of the mass function m(x)
defined by $m(x)=m_{0}M(x)$, and we have chosen units such that
$\hbar=2m_{0}=1$. The effective potential $V_{eff}$ contains,
apart from the given $V(x)$, the real ambiguity parameters
$\alpha$ and $\beta$ whose occurrences are typical in PDM settings
:
$$V_{eff}=V(x)+\frac{1}{2}(\beta+1)\frac{M''}{M^{2}}-\{\alpha
(\alpha+\beta+1)+\beta+1\}\frac{M'^{2}}{M^{3}}\eqno(2) $$ Suitable
physical
 choices of $\alpha$ and $\beta$
have been reported in the literature [4-25] but of particular
interest to us are the schemes of Zhu-Kroemer (ZK) [4]
$[\alpha=-1/2,\beta=0]$ and BenDaniel-Duke (BDD) [5]
$[\alpha=0,\beta=-1]$ which were shown [1] to be dual of each
other
for the free-particle case $V(x)=V_{0}$ that is independent of any choice of $M(x)$.\\

 \indent Substituting (2) in (1) and assuming for $V_{0}$ the form
$$V_{0}=\epsilon-\lambda(\lambda+1)q^{2} , \hspace{.5cm} \lambda,q \in \bf{R} \eqno(3)$$
 we can recast (1) to the standard constant-mass
Schr\"{o}dinger equation namely
$$(-\frac{d^{2}}{dx^{2}}+u)\psi=0 \eqno(4)$$ with the energy level
term missing. In (4), $u$ is given by
$$u=\Big[\frac{3}{4}-\{\alpha(\alpha+\beta+1)+\beta+1\}\Big]\frac{M'^{2}}{M^{2}}+
\frac{1}{2}\beta\frac{M''}{M}-\lambda(\lambda+1)Mq^{2}.\eqno(5)$$

However, Eq. (4) can also be regarded as the linearized partner of
the Riccati equation $$u=v^{2}+v'\eqno(6)$$ upon putting
$v=\frac{\psi^\prime}{\psi}$. The latter is the Cole-Hopf
transformation.\\

\indent A nonlinear connection such as the one given by (6), also
known as the Miura map, has an interesting implication. It
transfers a solution of the modified KdV equation
$$v_{t}=6v^{2}v'-v'''\eqno(7)
$$ into a solution of the KdV equation $$u_{t}=6uu'-u'''\eqno(8)
$$which is
straightforward to check.\\

The KdV equation has a very rich internal structure [26-27]. In
particular, it admits of a Lax representation $L_{t}=[B, L]$,
where $L=-\partial^2+u$ is a Schr\"{o}dinger-like operator and B
is given by $B=-4\partial^{3}_{x}+6u\partial_{x}+3u'$. One can
solve for L in the from $L(t)=S(t)L(0)S^{-1}(t)$ with $S_{t}=BS$.
The related eigenvalue problem then implies that the spectrum of L
is conserved and yields for the KdV an infinite chain of conserved
charges.\\

Noting that the KdV is invariant under the set of transformations
$$t \rightarrow t' ,\hspace{.5cm} x \rightarrow x'-6ct' ,
\hspace{.5cm} u \rightarrow u'+c \eqno(9)$$ where c is a constant,
the energy levels $\mu_{n}$ can be introduced in (4):
$$(-\frac{d^{2}}{dx^{2}}+u)\psi=\mu_{n}\psi .\eqno(10) $$
The manner of interplay between the PDM form (5) of $u$ for
specific choices of the parameters $\alpha, \beta$ and the initial
condition $u(x,0)=u_{0}$ used as inputs to solve for the KdV (as
is normally done in the inverse scattering problem) is our point
of
enquiry.\\

It can be proved that the discrete eigenvalues $\mu_{n}$ are
time-independent. For this we have to express the KdV in the
conserved form
$$u_{t}+(-3u^{2}+u_{xx})_{x}=0 \eqno(11)$$ and substitute u from (10) into it. We obtain
$$(\mu_{n})_{t} \psi^{2}+(\psi\phi_{x}-\psi_{x}\phi)_{x}=0\eqno(12) $$
where $\phi=\psi_{t}+\psi_{xxx}-3(u+\mu)\psi_{x}$. On integrating
(12) we find $(\mu_{n})_{t}=0$ where we have employed normalized
$\psi$ and considered vanishing asymptotic conditions for $\psi$
and its derivatives. The eigenvalues $\mu_{n}$ are determined
using for the potential the initial value $u_{0}$
that corresponds to a stationary soliton solution of the the KdV equation.\\

\indent In the context of (10), the Riccati equation (6) is
transformed to
$$u=v^2+v'+\mu\eqno(13) $$where v as a solution of the generalized MKdV
equation $$v_{t}=6(v^2+\mu)v'-v'''\eqno(14) $$ ensures that u
evolves according to
the KdV equation.\\

\indent For the 1-soliton and the 2-soliton solutions of the KdV,
the corresponding starting solutions $u_{0}$ along with  $\psi$, v
and
the eigenvalues $\mu_{n}$ (n=1, 2) are given by :\\

\hspace{2.25em} 1-soliton : $u_{0}^{(1)}=-2q^{2}sech^{2}qx$,
$$\psi_{1}=\frac{1}{\sqrt{2}}sechqx,\hspace{.5em} v^{(1)}=-q\tanh
qx,\hspace{.5em}\mu_{1}=-q^{2} \eqno(15)$$

\hspace{2.5em} 2-soliton : $u_{0}^{(2)}=-6q^{2}$sech$^{2}$qx ,
$$\psi_{2}^{(a)}=\frac{\sqrt{3}}{2}sech^{2}qx,\hspace{.5em}v_{2}^{(a)}=-2q
\tanh qx,\hspace{.5em}\mu_{2}^{(a)}=-4q^{2}$$
  and
$$\psi_{2}^{(b)}=\frac{\sqrt{3}}{2}sech
qx \tanh qx
,\hspace{.5em}v_{2}^{(b)}= q\frac{1-2\tanh^{2}qx}{\tanh qx},\hspace{.5em}\mu_{2}^{(b)}=-q^{2}.$$\\\\
where   note that for one discrete value of the Schr\"{o}dinger
equation (10), there exists a 1-soliton solution and vice-versa.
Similarly for the 2-soliton case. Here the $\psi$'s are normalized. \\

\indent The results in (15), which can also be extended to the
N-soliton case, have been obtained by solving the eigenvalue
problem for the Schr\"{o}dinger equation (10). The solutions
$u^{(1)}_{0}$ and $u^{(2)}_{0}$ act in (10) as the reflectionless
potentials. The inverse scattering method, which exploits this
reflectionless feature, determines the evolution of the scattering
parameters. Subsequently the Ge\'{l}fand-Levitan integral equation
is solved to
obtain the solution $u(x, t)$ of the KdV equation.\\

\indent Turning now to the PDM induced u given by (5), we
immediately recognize from (10) that for the choice of the mass
function $M(x)=$ sech$^{2}$qx, the ZK scheme yields the 1-soliton
results $u^{(1)}_{0},\psi_{1}(\mu=-q^{2})$ corresponding to
$\lambda=1,-2$ and the 2-soliton result
$u^{(2)}_{0},\psi^{(a)}_{2}(\mu=-4q^{2})$
and $\psi^{(b)}_{2}(\mu=-q^{2})$ corresponding to $\lambda=2,-3$.\\

On the other hand, the BDD scheme is consistent with the form\\

\hspace{3cm}$u=q^{2}(1-2$sech$^{2}qx),\hspace{.5cm}\lambda=1,-2\hspace{4.5cm}(16)$\\
for  $\psi_{1}(\mu=0)$ and \\

\hspace{3cm}$u=q^{2}(1-6$sech$^{2}qx),\hspace{.5cm}\lambda=2,-3\hspace{4.5cm}(17)$\vspace{.5cm}
for both the sets  $\psi^{(a)}_{2}(\mu=-3q^{2})$ and $\psi^{(b)}_{2}(\mu=0)$.\\

To interpret the above results, a few remarks on SUSY are in order
[28]. We first of all verify that not only (13) but also
$u=v^{2}-v'+\mu$ carries a solution of the generalized KdV (14)
into a solution of
the KdV.\\

Denoting $$V^{(\pm)}\equiv u^{\pm}-\mu=v^{2}\mp v'\eqno(18)$$we
notice that the combination $V^{(\pm)}$ can be identified as the
usual partner potentials of
SUSYQM.\\

\indent To examine the role of $V^{(\pm)}$ in the present context,
let there be a Hamiltonian $H_{1}$ with potential $V_{1}$ that is
asymptotically vanishing and having a set of n discrete
eigenvalues $\mu_{1}, \mu_{2}, ......, \mu_{n}$. If we define
$V^{+}=V_{1}-\mu_{n+1}$ then, in unbroken SUSY, we at once know
that the spectra of $V^{(+)}$ and those of $V^{(-)}$ are one to
one except that the latter has an additional $\mu=0$ state. In
other words, the eigenvalues of $V^{(-)}$ are
$\mu_{1}-\mu_{n+1},\mu_{2}-\mu_{n+1},.....,\mu_{n}-\mu_{n+1}$ and
0. This means that the Hamiltonian $H_{2}$ with potential $V_{2}$
defined by $V_{2}=V^{(-)}+\mu_{n+1}$ has (n+1) discrete
eigenvalues
$\mu_{1},\mu_{2},.....,\mu_{n},\mu_{n+1}$.\\

\indent Let us apply the above ideas to the simple case of
$V_{1}=0$ and generate the corresponding potential $V_{2}$ with a
single bound state with $\mu=-q^{2}$  [29]. We have
$V^{(+)}=v^{2}-v^{\prime}=q^{2}>0$: in other words,  $V^{(+)}$ has
no bound state at all. Solving we get $v=-q$tanh $qx$ (i.e. the
1-soliton result) which in turn gives $V^{(-)}=q^{2}$(1-
2sech$^{2}qx$) that supports a zero energy $(\mu=0)$ bound state
$\psi_{0}\sim sech qx :$
$$H_{-}\psi_{0}=\psi_{0}''+(v^{2}+v')\psi_{0}\eqno(19)$$
Thus $V_{2}=-2q^{2}$sech$^{2}qx$ has a single bound state.\\

\indent We immediately recognize $V_{2}$ and $V^{(-)}$ to be the
PDM potential u for the ZK and BDD schemes respectively
corresponding to the 1-soliton case. The same is true for the
2-soliton results with
$v$ matching with the 2-soliton solutions and $V^{(-)}$ emerging similar to (17).\\

\indent One-dimensional supersymmetric approach to PDM quantum
systems has been explored before in PDM scenarios. The partner
potentails were found to obey [22] the same PDM dependence but in
different potentials. The approach of this work is however
different in spirit from such a viewpoint in that we have sought
to establish a link between a hierarchy of reflectionless
potentials ( corresponding to the stationary soliton solutions of
the KdV ) with an arbitrary bound state spectrum and those of SUSY
in PDM models for suitable values of the ambiguity parameters. Our
starting potential pertaining to the free-particle case
$V(x)=V_{0}$ can be made to coincide with $V^{(+)}$ by choosing,
for example, $\epsilon=3q^{2}$ in the
1-soliton case and $\epsilon=12q^{2}$ in the 2-soliton case.\\

\indent Finally, we can extend our treatment to other special
cases of the effective potential $V_{eff}$ namely those of the
Bastard [30] and Li and Kuhn (redistributed) [31] Hamiltonians.
For the 1-soliton result of (15), $u$ for the Bastard scheme is $
u=-q^{2}(1+3$sech$^{2} qx)$ $(\mu=-2q^{2})$ while for the
2-soliton results given by (16), $u$ turns out to be
$-q^{2}(1+6$sech$^{2}qx)$ both for $\psi_{2}^{(a)}$ and
$\psi_{2}^{(b)}$, with an associated $\mu$-value of $\mu=-5q^{2}$
and $\mu=-2q^{2}$ respectively. However, in the Bastard model
$\lambda$ is non-integral. A non-integral  $\lambda$ also emerges
in the Li-Kuhn scheme where we find $ u=-2q^{2}$sech$^{2}qx
 (\mu=-q^2)$ for the 1-soliton solution  and $
u=-6q^2$sech$^{2}qx$ for both the 2-soliton solutions of
$\psi_{2}^{(a)}(\mu=-4q^{2})$ and $\psi_{2}^{(b)}
(\mu=-q^{2})$.\\\\
I thank P.S.Gorain for discussions.\\\\\\

{\large \bf References:} \vspace{.5cm}

[1] B. Bagchi, P. Gorain, C. Quesne,
R.Roychoudhury,Mod.Phys.Lett.{\bf A 19},
    2765 (2004).\\

[2]  See, for example, B. K. Bagchi, Supersymmetry in Quantum and Classical Mechanics, Chapman and Hall (Florida, 2000).\\

[3] O. von Roos, Phys. Rev. {\bf B 27}, 7547 (1983).\\

[4] Q.-G. Zhu and H. Kroemer, Phys. Rev {\bf B 27}, 3519 (1983).\\

[5] D.J. BenDaniel and C.B. Duke, Phys. Rev {\bf B 152}, 683 (1966).\\

[6] F. Arias de Saavedra, J. Boronat, A. Polls and A. Fabrocini, \\Phys. Rev {\bf B 50}, 4248 (1994).\\

[7] A. Puente, Ll. Serra and M .Casas, Z. Phys. {\bf D 31}, 283 (1994).\\

[8] J.-M.L\'{e}vy-Lebond, Phys. Rev {\bf A 52}, 1845 (1995).\\

[9] K.C. Yung and J.H. Yee, Phys. Rev. {\bf A 50}, 104 (1994).\\

[10] L. Chetouani, L. Dekar and T.F. Hammann, Phys. Rev. {\bf A 52}, 82 (1995).\\

[11] L. Dekar, L. Chetouani  and T.F. Hammann, J. Math. Phys. {\bf 39}, 2551 (1998).\\

[12] L. Dekar, L. Chetouani  and T.F. Hammann, Phys. Rev. {\bf A 59}, 107 (1999).\\

[13] A.R. Plastino, A. Puente, M. Casas, F. Garcias and A. Plastino, Rev. Mex. Fis. {\bf 46} 78 (2000).\\

[14] R. Ko\c{c}, M. Koca and E. K\"{o}rc\"{u}k, J. Phys. {\bf A 35}, L527 (2002).\\

[15] B. G\"{o}n\"{u}l, O. \"{O}zer, B. G\"{o}n\"{u}l and F.\"{U}zg\"{u}n, Mod. Phys. Lett {\bf A 17}, 2453 (2002).\\

[16] A.D. Alhaidari, Phys. Rev. {\bf A 66}, 042116 (2002).\\

[17] A. de Souza Dutra and C.A.S. Almeida, Phys. Lett. {\bf A 275}, 25 (2000).\\

[18] O.Mustafa and S.H.Mazharimousavi  J.Phys.{\bf A39}, 10537 (2006).\\

[19] J. Yu, S.-H. Dong and G.-H. Sun, Phys. Lett. {\bf A 322}, 290 (2004).\\

[20] B. Roy and P.Roy, J. Phys. {\bf A35}, 3961 (2002).\\

[21] V. Milanovi\'{c} and Z. Ikoni\'{c}, J. Phys. {\bf A32}, 7001 (1999).\\

[22] A.R. Plastino, A. Rigo, M. Casas, F. Garcias and A. Plastino, Phys. Rev. {\bf A 60} 4318 (1998).\\

[23] B. G\"{o}n\"{u}l,  B. G\"{o}n\"{u}l, D. Tuctu. and O. \"{O}zer, Mod. Phys. Lett {\bf A 17}, 2057 (2002).\\

[24] S Cruz y Czuz , J Negro , L M Nieto , Phys Lett {\bf A 369}, 400 (2007).\\

[25] J J Pena et al, Int. J. Quant. Chem {\bf 107}, 3039 (2007).\\

[26] R.M. Miura, SIAM Rev. {\bf 18}, 412 (1976).\\

[27] P. L. Bhatnagar, Nonlinear Waves in One-dimensional
dispersive
systems, Oxford University Press (Delhi, 1979).\\

[28] B. Bagchi, Int. J. Mod. Phys. {\bf A 5}, 1763 (1990).\\

[29] A.K. Grant and J.L. Rosner, J. Math. Phys {\bf 35}, 1 (1994).\\

[30] G. Bastard, Phys. Rev. {\bf B 24}, 5693 (1981).\\

[31] T.L. Li and K.J. Kuhn, Phys. Rev. {\bf B 47}, 12760 (1993).\\

\end{document}